\documentclass[10pt,twocolumn]{article}

\usepackage[
    letterpaper,
    top=0.75in,
    bottom=0.85in,
    left=0.70in,
    right=0.70in,
    columnsep=0.25in
]{geometry}

\usepackage[T1]{fontenc}
\usepackage[utf8]{inputenc}
\usepackage{newtxtext}
\usepackage{newtxmath}

\usepackage{microtype}
\usepackage{graphicx}
\usepackage{booktabs}
\usepackage{multirow}
\usepackage{amsmath}
\usepackage{mathtools}
\usepackage{bm}
\usepackage{xcolor}
\usepackage{enumitem}
\usepackage{caption}
\usepackage{subcaption}
\usepackage{wasysym}
\usepackage{pdfpages}

\usepackage{algorithm}
\usepackage{algorithmic}

\usepackage{tabularx}
\usepackage{array}

\usepackage[numbers,sort&compress]{natbib}

\usepackage{xurl}
\usepackage[
    colorlinks=true,
    linkcolor=blue,
    citecolor=blue,
    urlcolor=blue
]{hyperref}
\usepackage{titlesec}

\titleformat{\section}
  {\large\bfseries}
  {\thesection}
  {0.5em}
  {}

\titleformat{\subsection}
  {\normalsize\bfseries}
  {\thesubsection}
  {0.5em}
  {}

\titleformat{\subsubsection}
  {\normalsize\bfseries}
  {\thesubsubsection}
  {0.5em}
  {}

\titlespacing*{\section}
  {0pt}{2.0ex plus 0.5ex minus 0.2ex}{0.8ex}

\titlespacing*{\subsection}
  {0pt}{1.5ex plus 0.3ex minus 0.2ex}{0.5ex}

\titlespacing*{\subsubsection}
  {0pt}{1.2ex plus 0.3ex minus 0.2ex}{0.4ex}

\setlist{
    topsep=2pt,
    itemsep=1pt,
    parsep=0pt,
    leftmargin=*
}

\title{
    \vspace{-0.5cm}
    \textbf{\LARGE Where Does AI Innovation Go? \\
Measuring Research Attention Imbalance in AI Music}
    \vspace{-0.2cm}
}

\author{
Qian Liang$^{1,\clubsuit}$,
Yanzhen Ning$^{2,\clubsuit}$,
Fengyuan Zhang$^{3}$,
Bo Dai$^{4}$,
Ningbo Cheng$^{1,\blacklozenge}$\\[6pt]
$^{1}$State Key Laboratory of Multimodal Artificial Intelligence System,\\
Institute of Automation, Chinese Academy of Sciences\\
$^{2}$School of Computer Science and Technology,\\
University of Chinese Academy of Sciences\\
$^{3}$School of Computer Science and Engineering, Northeastern University\\
$^{4}$Composition Department, Central Conservatory of Music\\[5pt]
\texttt{\{qian.liang, ningbo.cheng\}@ia.ac.cn}\\
\texttt{ningyanzhen23@mails.ucas.ac.cn},
\texttt{20236198@stu.neu.edu.cn},
\texttt{dbo@ccom.edu.cn}
}

\date{}

\begin{document}

\maketitle
\begingroup
\makeatletter
\renewcommand{\thefootnote}{}
\renewcommand{\@makefntext}[1]{\noindent #1}
\makeatother

\footnotetext{
\footnotesize
$\spadesuit$ These authors contributed equally to this work.\\[-1pt]
$\blacklozenge$ Corresponding author: \texttt{ningbo.cheng@ia.ac.cn}.
}
\endgroup

\renewenvironment{abstract}
{
    \begin{center}
        \vspace{-0.5em}
        {\large\sffamily\bfseries Abstract}
        \vspace{0.2em}
    \end{center}
    \small\sffamily
    \noindent
}
{
    \par
    \normalfont\normalsize
    \vspace{0.6em}
}
\begin{abstract}
The rapid growth of artificial intelligence (AI) in music has expanded research from generation and information retrieval to education, health, and governance. Yet this growth does not necessarily imply balanced research attention. Where is research attention directed across diverse music tasks, and how can such imbalance be systematically measured? Existing studies examine AI music from separate technical, application-specific, or bibliometric perspectives, but lack a systematic framework for measuring field-level imbalance. To address this gap, we analyze 6,839 AI music publications from 2015 to April 2026 using a joint taxonomy of 12 application categories and 11 technical method families. We propose the Research Attention Profile, comprising four indicators of technical investment, method allocation, methodological diversity, and frontier-method adoption lag. Results show that technical support is concentrated in scalable, content-oriented tasks, while education, health, and governance remain under-supported. Generation adopts frontier methods after only 0.33 years on average, compared with 4.33 years for education and 5.00 years for health. These findings reveal uneven methodological development and support a more socially responsive AI music research agenda.
\end{abstract}
\begin{figure}[ht]
    \centering
    \includegraphics[width=1.0 \columnwidth]{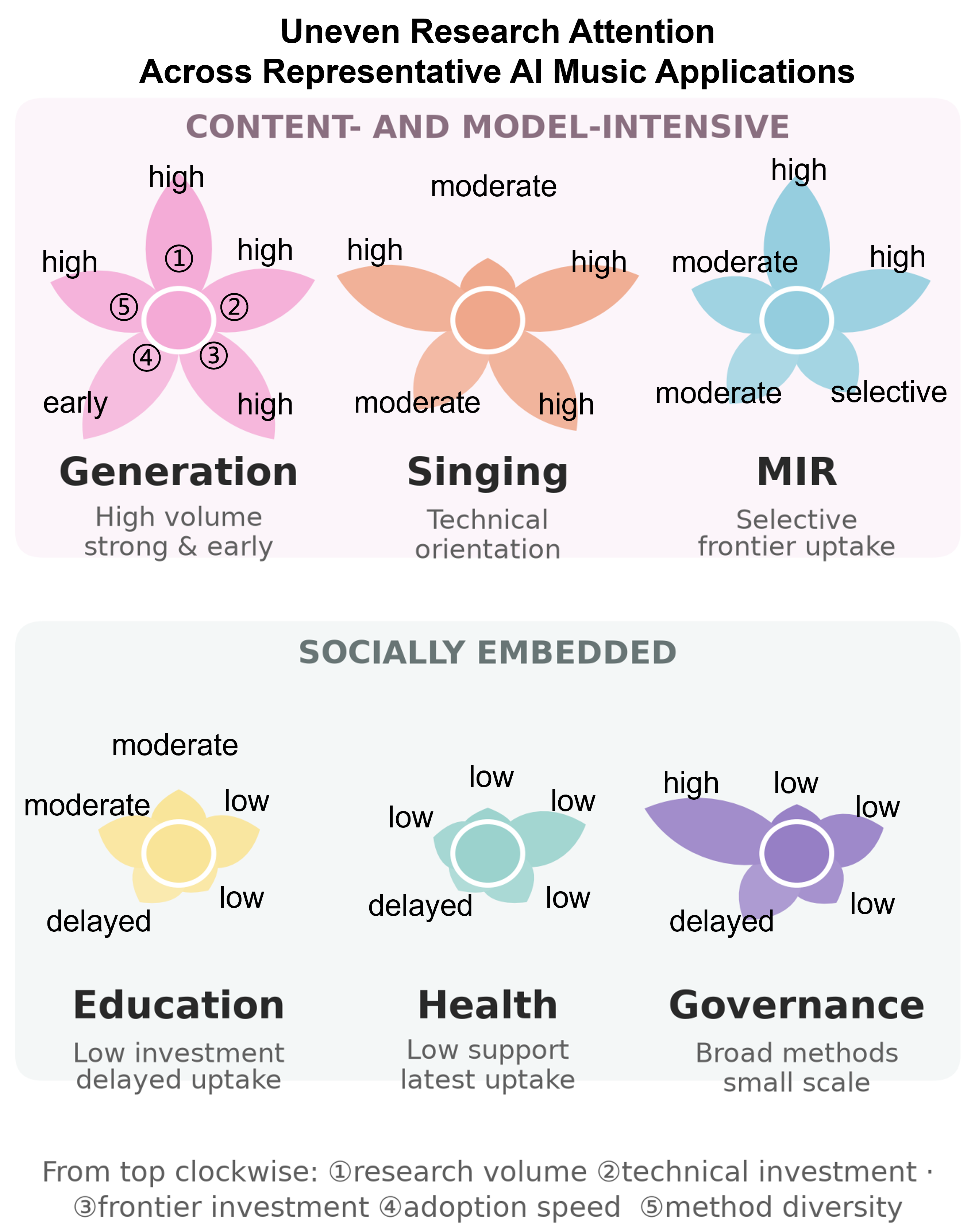}
    \caption{Research attention imbalance across representative AI music applications. Petal length indicates research volume, technical and frontier-method investment, adoption speed, and methodological diversity.}
    \label{fig:flower}
\end{figure}
\section{Introduction}

The rapid development of Artificial Intelligence (AI) is increasingly embedded in cultural, creative, educational, health-related, and other practices~\cite{stanford2025AIreport}. Music provides a particularly informative case at the intersection of humanistic expression, creativity, social participation~\cite{denora2000music}, and AI methodology. AI music research now spans generation~\cite{Ji_2023_05}, understanding~\cite{le2025natural},  recommendation~\cite{schedl2019deep}, education~\cite{chen2024usage}, health~\cite{shen2024first}, and human-AI creative assistance~\cite{mahmud2023study}, among others. 

However, rapid growth does not necessarily imply a balanced allocation of research attention. In this paper, we use \textit{research attention} to refer to both the application problems that receive scholarly focus and the methodological support allocated to them. Some tasks are supported by abundant data, established benchmarks, and clearly defined computational objectives. Other areas, including health, copyright, ethics, and governance, have direct social consequences but involve stronger human, clinical, legal, and institutional contexts. This raises two questions:

\textbf{RQ1.} \textit{Does AI music research reflect balanced methodological investment across societal needs, or unequal attention toward technically tractable, model-driven tasks?}

\textbf{RQ2.} \textit{If such imbalance exists, how can it be systematically measured?}

Answering these questions requires analyzing AI music as an evolving research ecosystem in which technical methods, application tasks, and social needs are jointly considered. Existing studies have examined this field from separate perspectives, including technical development~\cite{briot2020deep}, application tasks~\cite{Ji_2023_05}, social values~\cite{gwerevende2023safeguarding}, and bibliometric mapping~\cite{gomez2025beyond}. However, these perspectives rarely connect \textit{what} music-related problems are studied with \textit{how} AI methods are used to address them, and with \textit{which} broader needs may be prioritized or under-supported. Consequently, existing work provides limited tools to determine whether methodological support is proportionally, broadly, and promptly distributed across application tasks, especially those with direct implications for health, rights, and governance.

To address this gap, we develop an analytical framework termed the \textit{Research Attention Profile}, linking application tasks, technical methods, and publication years to characterize the distribution and temporal evolution of research attention in AI music. As shown in Fig.~\ref{fig:flower}, content- and model-intensive
tasks receive stronger and earlier technical support than social related applications. Our contributions are as follows:
\begin{itemize}
    \item We construct a large-scale corpus which contains $6,839$ AI music research publications from 2015 to 2026 and developed a joint application-technology taxonomy that characterizes both the task space and the AI method space of this domain.
    \item We introduce a computational measurement method, including adoption intensity, adoption timing, technological diversity, and coupling concentration, to quantify uneven method diffusion across AI music tasks.
    \item We introduce the Research Attention Profile comprising four key indicators, Technical Investment Residual, Task-Method Investment Residual, Normalized Methodological Diversity and Frontier Method Adoption Lag to systemically measure complementary differences in the amount, allocation, structure, and timing of technical support. 
    \item We provide empirical evidence of uneven development in AI music research, identifying the selective concentration of technical and frontier-method investment in content-production tasks and the comparatively weaker and slower methodological support received by education, health, and governance (as shown in Fig.~\ref{fig:flower}).
\end{itemize}

\section{Related Works}
Since the early demonstrations of computer-generated music and singing at Bell Labs~\cite{mathews1963digital}, computational approaches to music creation have evolved for nearly seven decades. With the rapid development of AI technologies, AI music has entered a new phase of expansion, researchers have examined this field from different perspectives, including model-driven, task-driven, social and cultural, and bibliometric mapping studies. 

Model-driven studies mainly focus on the technical development of AI methods for music. A substantial body of work has established the methodological foundations of this field, ranging from rule-based systems~\cite{ebciouglu1988expert}, statistical models~\cite{conklin1995multiple} to deep learning architectures~\cite{briot2020deep}, such as convolution neural network (CNN) based models~\cite{choi2017convolutional}, sequence models~\cite{eck2002finding}, generative adversarial networks~\cite{dong2018musegan}, Transformers~\cite{huang2018music}, diffusion models~\cite{huang2023noise2music}, spiking neural network (SNN) based models~\cite{liang2026spiking} and recent foundation-models~\cite{Ma2024FoundationMF}, especially those based on large language models~\cite{wang2025notagen}. These studies are valuable for understanding how AI techniques enter music research and how technical paradigms change over time. However, they provide limited insight into how these models are distributed across different music application tasks. This leaves unclear the task-oriented structure of AI music research.

Task-driven studies examine AI music through specific application areas. Existing reviews have discussed music generation~\cite{Ji_2023_05}, music information retrieval~\cite{lerch2022introduction,le2025natural}, music recommendation~\cite{schedl2019deep}, music emotion recognition~\cite{liyanarachchi2025survey}, music education~\cite{chen2024usage}, music and health~\cite{shen2024first}, and human-AI creative assistance~\cite{mahmud2023study}. These studies clarify the problem definitions, datasets, evaluation protocols, and domain-specific concerns of individual tasks. However, because each application area is usually reviewed within its own methodological tradition, this perspective remains fragmented, and leaves unresolved how these tasks differ in their absorption of AI method families over time.

Human-centered, social, and cultural perspectives approach music highlight the broader value of music for human expression~\cite{denora2000music,thompson2023psychological}, creativity~\cite{deliege2006musical}, well-being~\cite{macdonald2013music}, cultural memory~\cite{bennett2016popular}, and social participation and transmission~\cite{gwerevende2023safeguarding}. Although recent discussions on AI music have addressed issues such as copyright~\cite{deng2023computational}, authorship~\cite{ylikallio2025musical}, labor~\cite{atanasovski2024artificial}, fairness~\cite{wang2024fairness},  and ethical governance~\cite{herington2026musicians}, while international policy frameworks emphasize cultural diversity~\cite{unesco2005convention}. However, this line of work is often distributed across humanities, social sciences, law, ethics, and cultural policy. It therefore rarely examines whether socially meaningful music applications receive comparable methodological investment within the AI music research ecosystem.

Bibliometric and science-mapping studies provide another perspective for examining the development of AI music and related music technology fields. Existing work has reported western-centered phenomenon by analyzing the authorship structure of the ISMIR community~\cite{gomez2025beyond} and broader AI-in-music research across creation, performance, and education~\cite{tong2026artificial}. These studies reveal field-level trends, but remain largely descriptive. They rarely provide task-level quantitative indicators linking music applications, AI method families, and temporal adoption patterns, making it difficult to diagnose research attention inequality in AI music.

\section{Data and Taxonomy}
\subsection{AI Music Corpus Collection}

We construct a large-scale corpus of AI music research articles published between 2015 and April 2026. Candidate papers were retrieved from three complementary sources: Semantic Scholar, the International Society for Music Information Retrieval (ISMIR), and arXiv. Semantic Scholar provides multidisciplinary coverage across journals and conferences, ISMIR captures a central body of specialized music information retrieval research, and arXiv includes rapidly developing work that may not yet have appeared in formal publication venues. Using a set of AI- and music-related keywords, we retrieved candidate records from all three sources; the full keyword list is provided in Table~S1 of the supplementary material. We then performed deduplication and preprocessing to remove repeated records across sources, normalize publication years and source names, and exclude papers not directly related to AI music research. The final corpus contains 6,839 papers, including 2,546 from Semantic Scholar (37.2\%), 1,104 from ISMIR (16.2\%), and 3,189 from arXiv (46.6\%). Each record contains bibliographic metadata, including title, abstract, publication year, source, and available affiliation or country information.

\subsection{Taxonomy Construction}
To characterize both what AI music research studies and how it technically approaches these problems, we further develop a joint taxonomy through a hybrid process that combines corpus-based keyword clustering, LLM-assisted category generation, and human curation. Details of the taxonomy construction procedure are provided in the S1.1 of supplementary material. The final taxonomy is shown in Figure~\ref{fig:taxonomy}, the joint taxonomy covers both application-task and technical-method dimensions: 
\begin{figure}[h]
\centering
\includegraphics[width=1.0\columnwidth]{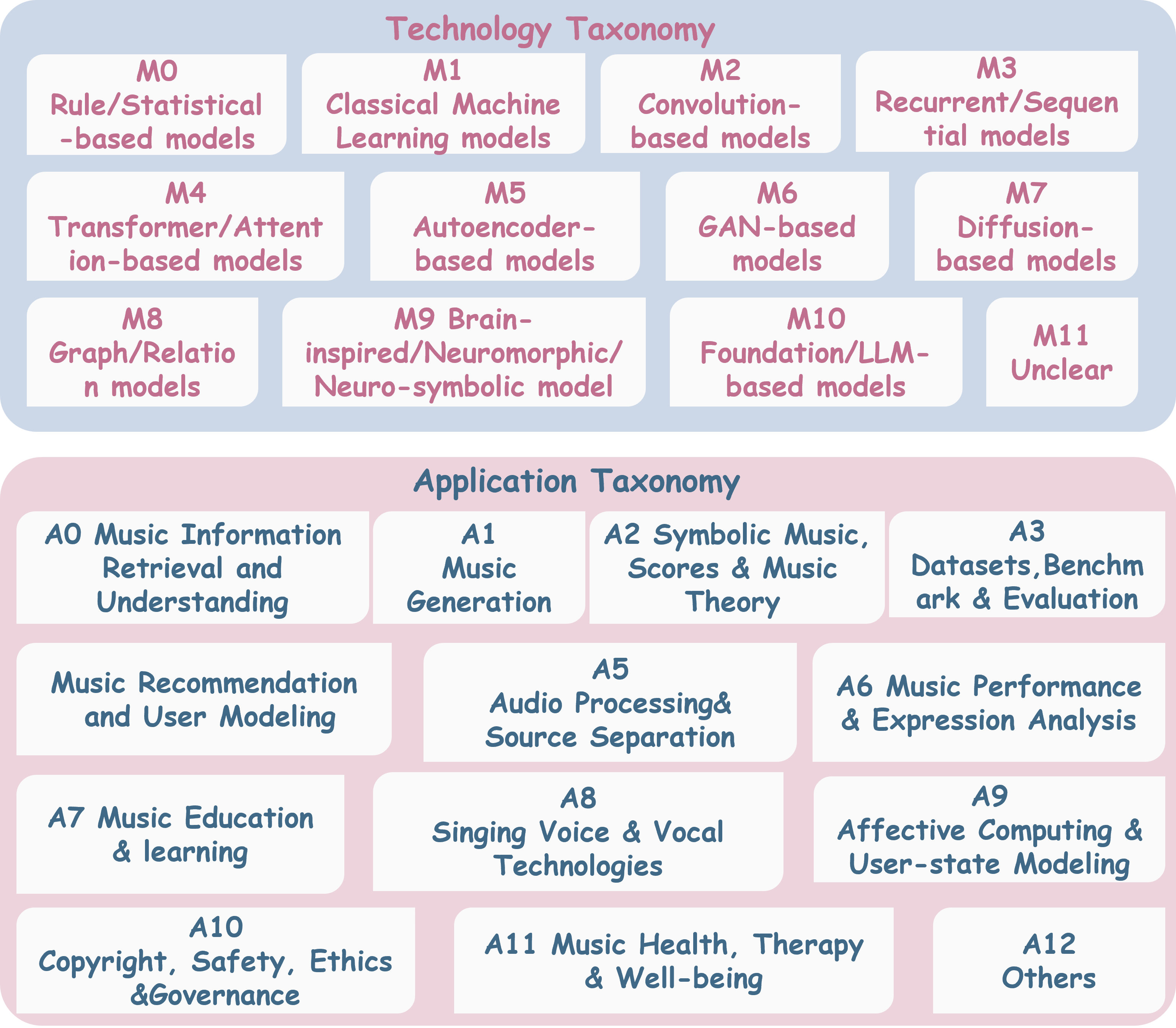} 
\caption{Joint taxonomy of AI music research. The upper panel presents the 12 primary technical-method categories, while the lower panel presents the 13 primary application categories.}
\label{fig:taxonomy}
\end{figure}
The application taxonomy includes 12 primary categories and 77 subcategories, covering music information retrieval, music generation, music recommendation, music education, music health, among others. The subcategories further specify the scope of each primary category, with details provided in the supplementary material.

The technical taxonomy includes 12 primary categories and 67 subcategories, covering mainstream AI model families developed over the past decades, such as CNN-based models, RNN-based models, transformer-based, foundation models(e.g. LLM, agent-based models), among others. The subcategories further specify architectures and learning paradigms within each primary category, with details provided in the supplementary material. 

This joint taxonomy links music application tasks with AI method families, providing the structural basis for measuring how methodological investment is distributed across the AI music research ecosystem.

\subsection{Taxonomy Mapping}
After the taxonomy construction, we assigned each paper in the corpus to single primary labels in application and technical dimensions. Since each paper addresses at least one music-related problem, application labels were assigned to the full corpus based on each paper's title, abstract, and keywords. For the technical dimension, we first examined whether a paper explicitly proposed or applied an AI method, and then assigned a primary technical label only to papers identified as technical.
\begin{algorithm}[h]
\caption{Primary Label Mapping and Validation}
\label{alg:primary_mapping}
\begin{algorithmic}[width=0.7\textwidth]
\REQUIRE Corpus $\mathcal{D}$; taxonomy $\mathcal{C}$; LLM classifier $f_{\theta}$; thresholds $\tau_{\kappa}, \tau_{F1}$
\ENSURE Primary label $y_i$ for each paper $p_i \in \mathcal{D}$

\STATE Assign initial labels $\hat{y}_i=f_{\theta}(p_i,\mathcal{C})$ for all $p_i \in \mathcal{D}$
\STATE Draw a stratified validation sample $\mathcal{S}$ based on $\hat{y}$
\STATE Two annotators independently label $\mathcal{S}$ and compute Cohen's $\kappa$
\IF{$\kappa < \tau_{\kappa}$}
    \STATE Refine annotation guidelines and repeat validation
\ENDIF
\STATE Resolve disagreements to obtain reference labels $y_{\mathcal{S}}^{ref}$
\STATE Evaluate $\hat{y}_{\mathcal{S}}$ against $y_{\mathcal{S}}^{ref}$ using Accuracy, and F1
\IF{$F1 \geq \tau_{F1}$}
    \STATE Accept $\hat{y}_i$ as final labels for all $p_i \in \mathcal{D}$
\ELSE
    \STATE Refine prompts or rules and repeat mapping
\ENDIF
\end{algorithmic}
\end{algorithm}

We employed LLM-assisted classification combined with stratified sampling and human validation for taxonomy mapping. As shown in Algorithm~\ref{alg:primary_mapping}, the procedure consists of four steps: initial LLM labeling, stratified sampling, independent human annotation, and validation against reconciled human labels. This protocol was applied to three classification tasks: application mapping, technical-relevance filtering, and technical-method mapping. For each task, two annotators independently labeled the sampled papers, and inter-annotator agreement was measured using Cohen's Kappa. The reconciled human labels were then used as reference labels to evaluate the LLM outputs using accuracy and macro-F1. We set the acceptance thresholds to $\tau_{\kappa}=0.80$, $\tau_{\mathrm{Acc}}=0.90$, and $\tau_{F1}=0.90$. As shown in Table.~\ref{tab:mapping_validation}, we therefore accepted the LLM-generated classification results and used them in the subsequent analysis.

\begin{table}[h]
\centering
\caption{Validation results of taxonomy mapping.}
\label{tab:mapping_validation}
\begin{tabular}[width=0.6 \textwidth]{lcccc}
\toprule
Task  & Sample Size & $\kappa$ & Accuracy & macro-F1 \\
\midrule
Tech-filtering  & 500 & 0.98 & 96.60\% & 94.16\% \\
Tech-mapping  & 403 & 0.93 & 93.55\% & 93.91\%\\
App-mapping  & 500 & 0.93 & 94.60\% & 94.18\%\\
\bottomrule
\end{tabular}
\end{table}

\section{Research Attention Profile}
We propose a measurement framework, termed the \textit{Research Attention Profile}, including the following four indicators to describe different aspects of research attention from task and technology dimensions. Let $\mathcal{D}$ denote the full corpus and $\mathcal{D}^{\mathrm{tech}} \subset \mathcal{D}$ denote the subset of technical papers.
\begin{itemize}
    \item \textbf{Technical Investment Residual (TIR)} is defined to measure whether an application task receives more or less technical investment than expected after controlling for its publication volume and the annual proportion of technical papers. For application task $a$ in year $y$, the expected number of technical papers is defined as
\begin{equation}
    E_{a,y}^{\mathrm{tech}}
    =
    n_{a,y}\frac{Q_y}{N_y},
\end{equation}
where $n_{a,y}$ and $N_y$ denote, the number of papers associated with task $a$ and the total number of papers in year $y$ within the full corpus $\mathcal{D}$, respectively. Similarly, $q_{a,y}$ and $Q_y$ denote the number of technical papers associated with task $a$ and the total number of technical papers in year $y$ within the technical subset $\mathcal{D}^{\mathrm{tech}}$.  The TIR is then calculated as
\begin{equation}
    \mathrm{TIR}_a
    =
    \frac{\sum_y \left(q_{a,y}-E_{a,y}^{\mathrm{tech}}\right)}
    {\sum_y E_{a,y}^{\mathrm{tech}}+\epsilon},
\end{equation}
where $q_{a,y}$ is the observed number of technical papers for task $a$ in year $y$, and $\epsilon$ is a small constant added for numerical stability. A positive $\mathrm{TIR}_a$ indicates higher technical investment, whereas a negative value indicates relative underinvestment.
    \item \textbf{Task-Method Investment Residual (TMIR)} measures whether a specific technical method receives more or less investment within an application task than expected from the task's overall research attention. This indicator combines the application distribution of the full corpus $\mathcal{D}$ with the method distribution of the method-mapped technical subset $\mathcal{D}^{\mathrm{tech}}$. Let
$o_{a,m,y}$ denote the observed number of papers linking application task $a$ with technical method $m$ in year $y$. The expected number of papers is defined as
\begin{equation}
    E_{a,m,y}^{\mathrm{method}}
    =
    n_{a,y}\frac{q_{m,y}}{N_y},
\end{equation}
where $n_{a,y}$ is the number of all papers associated with task $a$, $N_y$ is the total number of papers in year $y$, and $q_{m,y}$ is the number of method-mapped technical papers using method $m$ in that year. The TMIR for task $a$ and method $m$ over the study period is calculated as
\begin{equation}
    \mathrm{TMIR}_{a,m}
    =
    \frac{
    \sum_y\left(
    o_{a,m,y}-E_{a,m,y}^{\mathrm{method}}
    \right)
    }{
    \sum_y E_{a,m,y}^{\mathrm{method}}
    }.
\end{equation}
A positive value indicates that method $m$ is allocated to task $a$ more frequently than expected from the task's overall publication share, whereas a negative value indicates lower investment. A value of zero indicates that the observed method investment is proportional to the task's overall research attention. 

    \item \textbf{Normalized Methodological Diversity (NMD)} measures how evenly technical methods are distributed within an application task. This indicator is calculated on the technical subset $\mathcal{D}^{\mathrm{tech}}$. For application task $a$ in year $y$, the proportion of papers using method $m$ is defined as
    \begin{equation}
        p_{a,m,y} = \frac{q_{a,m,y}}{q_{a,y}},
    \end{equation}
    where $q_{a,m,y}$ is the number of technical papers linking task $a$ with method $m$ in year $y$, and $q_{a,y}=\sum_{m\in\mathcal{M}}q_{a,m,y}$ is the total number of technical papers associated with task $a$ in that year. Here, $\mathcal{M}$ denotes the set of method (model families) categories. The normalized methodological diversity is calculated as
    \begin{equation}
        \mathrm{NMD}_{a,y}=\frac{ -\sum_{m\in\mathcal{M}} p_{a,m,y}\log p_{a,m,y} }
        {\log |\mathcal{M}|}.
    \end{equation}
    The value ranges from 0 to 1. A value close to 0 indicates that the task is concentrated on one or a few methods, whereas a value close to 1 indicates a more even distribution across method categories. The indicator can also be calculated for a selected period by aggregating $q_{a,m,y}$ over the corresponding years before computing the method proportions.
    
    \item \textbf{Frontier Method Adoption Lag (FMAL)} describes how quickly an application task adopts a frontier method after the method enters AI music research. This indicator is calculated on the technical subset $\mathcal{D}^{\mathrm{tech}}$. Let $\mathcal{F}\subseteq\mathcal{M}$ denote the set of frontier methods. For method $m\in\mathcal{F}$, let $t_m^{0}$ denote its first adoption year in the technical corpus and $t_{a,m}$ denote its first adoption year within application task $a$. The adoption lag is defined as
    \begin{equation}
        \mathrm{FMAL}_{a,m}
        =
        t_{a,m}-t_m^{0}.
    \end{equation}
    A value of 0 indicates that task $a$ adopts method $m$ in the same year that it enters AI music research, whereas a larger positive value indicates a longer delay. For each task, the average adoption lag across the frontier methods it adopts is calculated as
    \begin{equation}
        \mathrm{FMAL}_{a}
        =
        \frac{1}{|\mathcal{F}_a|}
        \sum_{m\in\mathcal{F}_a}
        \mathrm{FMAL}_{a,m},
    \end{equation}
    where $\mathcal{F}_a$ is the set of frontier methods adopted by task $a$. Methods not adopted by the end of the study period are reported separately rather than assigned an arbitrary lag.

\end{itemize}

\begin{figure*}[ht]
\centering
\includegraphics[width=1.0\textwidth]{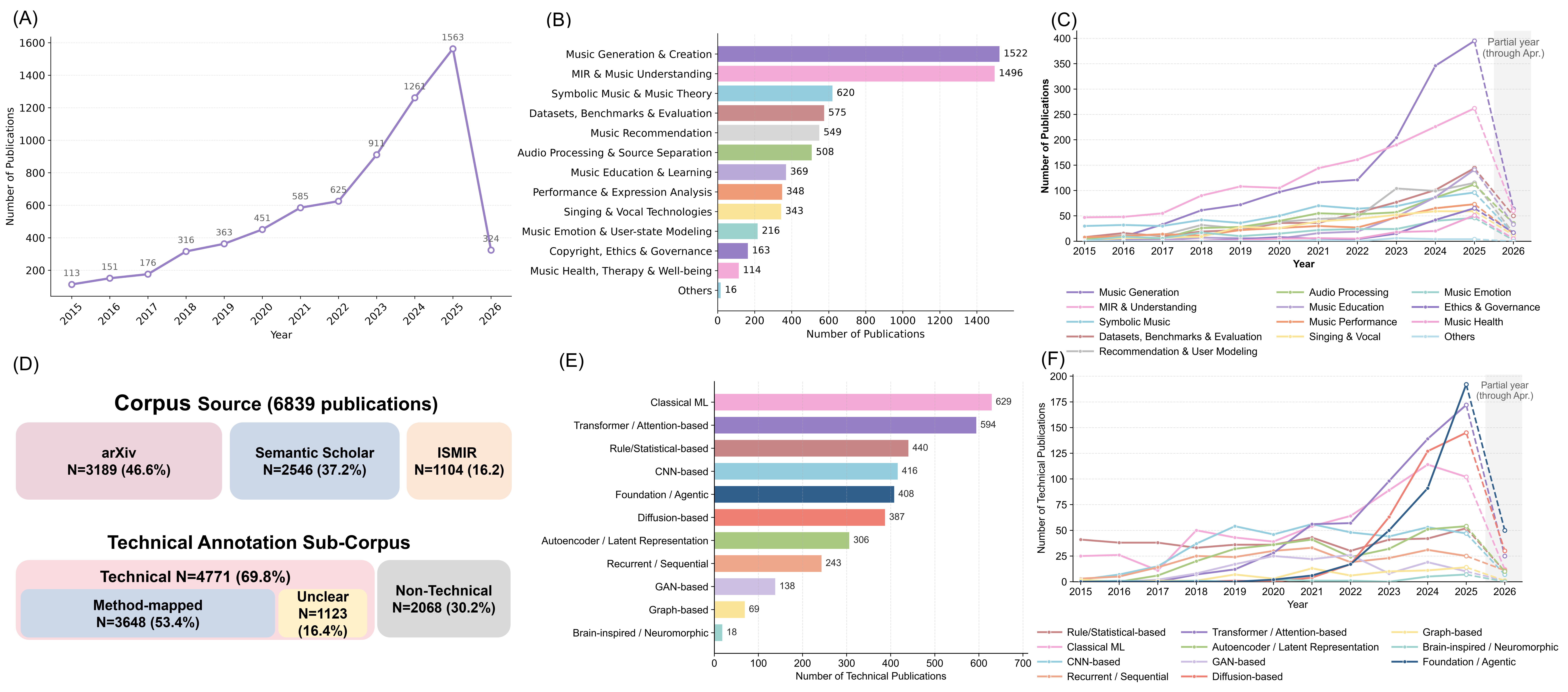} 
\caption{Overview of the AI music research landscape. 
(A) Annual publication trends from 2015 to 2026, showing the overall growth of research attention. 
(B) Distribution of application tasks across the AI music literature. 
(C) Annual publication trends of the application categories. 
(D) Corpus profile, including publication volume, temporal coverage, taxonomy size, source composition, and technical-annotation status. . 
(E) Distribution of technical methods among technical publications.
(F) Annual publication trends of the technical categories. }
\label{fig:overview}
\end{figure*}
\section{Results}
\subsection{Overview}
Based on the corpus and taxonomy developed in this study, we first present an overview of the AI music research landscape. Figure~\ref{fig:overview}A shows a sustained increase in publication volume, from 113 papers in 2015 to 1,563 in 2025. The 2026 count is incomplete because the corpus was collected only through April. 

Figure~\ref{fig:overview}D summarizes the corpus profile. The final dataset contains 6,839 publications, of which 3,648 (53.3\%) were explicitly mapped to one of the 11 technical-method families. The remaining papers were identified as non-technical ($n=2{,}068$) or unclear from their titles and abstracts ($n=1{,}123$).

Figure~\ref{fig:overview}B presents the application distribution across the full corpus. Music generation and creation receives the largest amount of research attention ($n=1{,}522$), closely followed by music information retrieval and understanding ($n=1{,}496$), while governance ($n=163$) and health-related applications ($n=114$) remain comparatively limited. Figure~\ref{fig:overview}C further shows that MIR dominated the earlier period, whereas music generation expanded rapidly and surpassed MIR after 2023.

Accordingly, Figures~\ref{fig:overview}E and~\ref{fig:overview}F report results only for the method-mapped technical subset. Classical machine learning ($n=629$) and Transformer-based models ($n=594$) are the two largest method families. The annual trends further reveal a shift from rule-based, statistical, and classical machine-learning approaches toward Transformer, diffusion, and foundation-model-based methods in recent years.
\begin{figure*}[ht]
    \centering
    \includegraphics[width=1.0\linewidth]{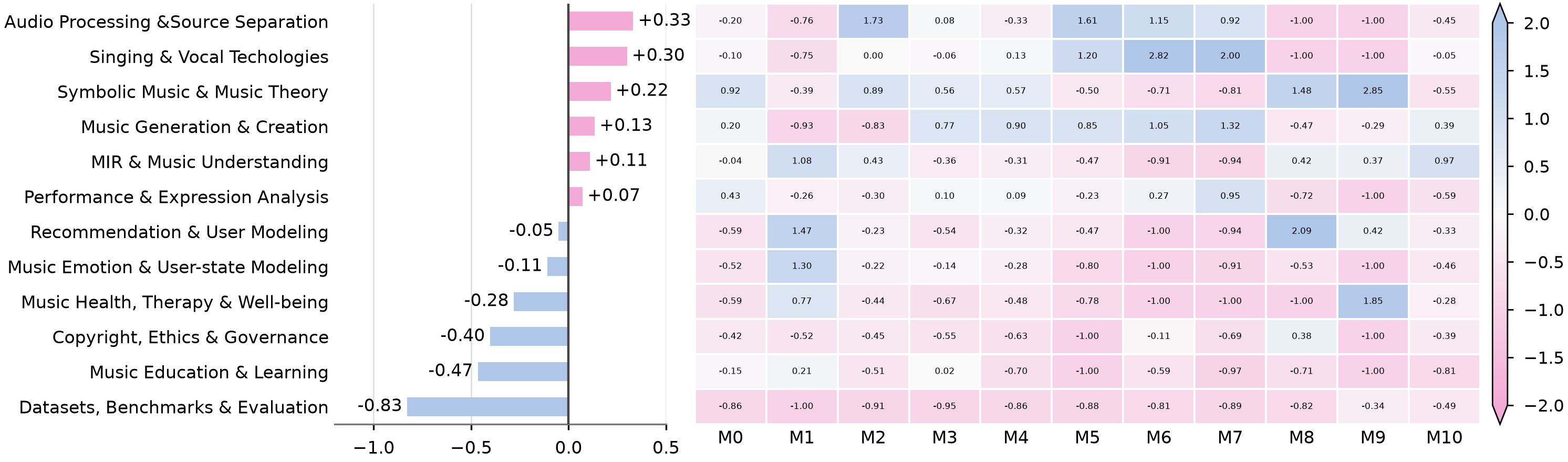}
    \caption{Technical investment and method allocation across AI music applications. The left bar panel indicates Technical Investment Residuals (TIR) across application categories. The right heatmap draws Task-Method Investment Residuals (TMIR) across all model families. Positive values indicate above-expected investment, while negative values indicate below-expected investment.}
    \label{fig:TIR_TMIR}
\end{figure*}
\subsection{Technical Investment and Allocation Across Tasks}
Figure.~\ref{fig:TIR_TMIR}A shows marked differences in technical investment across applications. Audio processing and source separation had the highest TIR ($0.331$). Generation and MIR also received more technical investment than expected, whereas education ($-0.466$), governance ($-0.404$), health ($-0.282$), and datasets and benchmarks ($-0.829$) received less.

\begin{figure}[h]
    \centering
    \includegraphics[width=1.0\columnwidth]{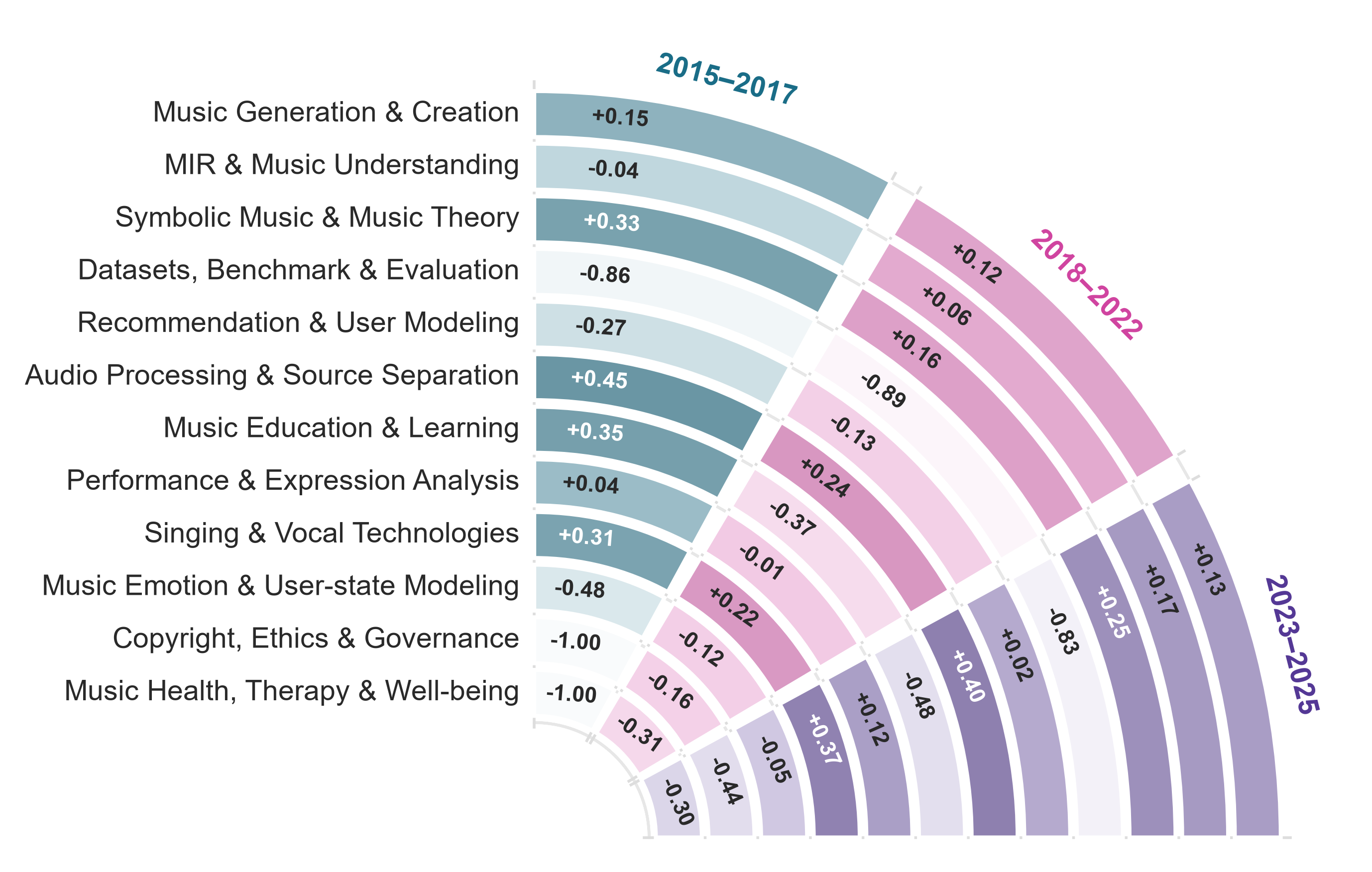}
    \caption{Temporal evolution of Technical Investment Residuals (TIR) across application categories in 2015--2017, 2018--2022, and 2023--2025. Positive values indicate above-expected technical investment, while negative values indicate low expected investment.}
    \label{fig:TIR_fanheat}
\end{figure}

We further calculated TIR for three periods: 2015--2017, 2018--2022, and 2023--2025. As shown in Figure~\ref{fig:TIR_fanheat}, generation and audio processing maintained positive residuals across all three periods. MIR and singing and voice technologies showed higher TIR values in the most recent period, while recommendation moved closer to the expected level. In contrast, education shifted from a positive residual in 2015--2017 to negative residuals in the two later periods. Governance, health, and datasets and benchmarks also remained below expectation in recent years.
Fig.~\ref{fig:TIR_TMIR}B shows the allocation of current frontier methods across tasks. Generation received substantially more Transformer- and diffusion-based investment than expected ($\mathrm{TMIR}=0.902$ and $1.317$), which is consistent with the strong technical orientation of adjacent content-production tasks, including singing and audio processing. The foundation-model investment in MIR reached nearly twice the expected level ($\mathrm{TMIR}=0.973$), whereas education, health, and governance remained below expectation across all three frontier-method families.

\subsection{Diversity and Timing of Technical Support}
\begin{figure}[h]
    \centering
    \includegraphics[width=0.5\textwidth]{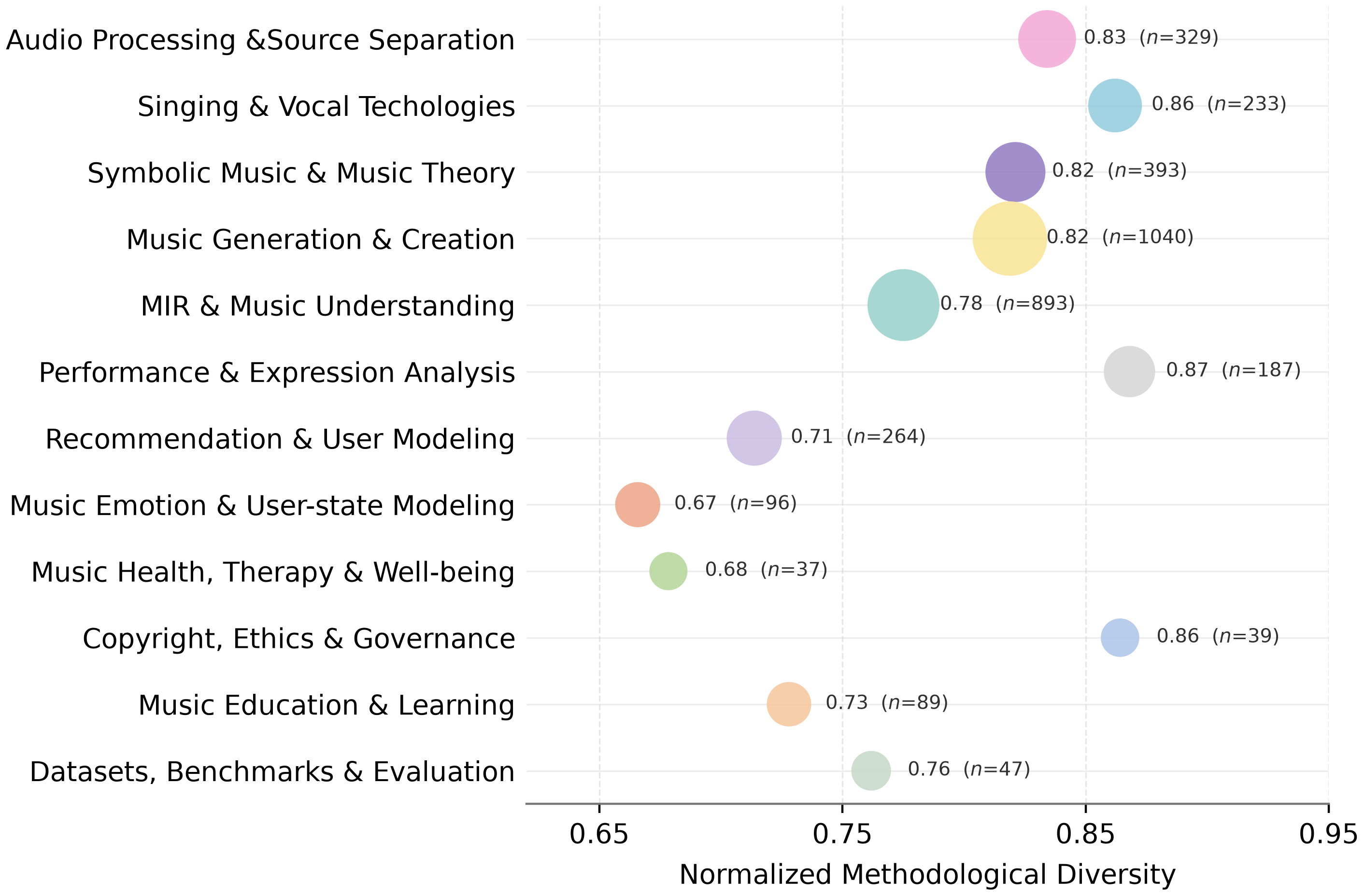}
    \caption{Normalized Methodological Diversity (NMD) across AI music application categories. Higher values indicate a more diverse distribution of technical methods within a task.}
    \label{fig:NMD}
\end{figure}

Fig.\ref{fig:NMD} shows the technical diversity across applications with publication amount. Generation got the most publications ($n=1040$) with relatively high diversity ($\mathrm{NMD}=0.819$), whereas education and health showed lower values of $0.728$ and $0.678$, respectively. Governance exhibited high diversity ($0.864$), but this estimate was based on only 39 mapped papers, indicating dispersion within a small technical literature rather than broad methodological support.

\begin{figure}[h]
    \centering
    \includegraphics[width=0.5\textwidth]{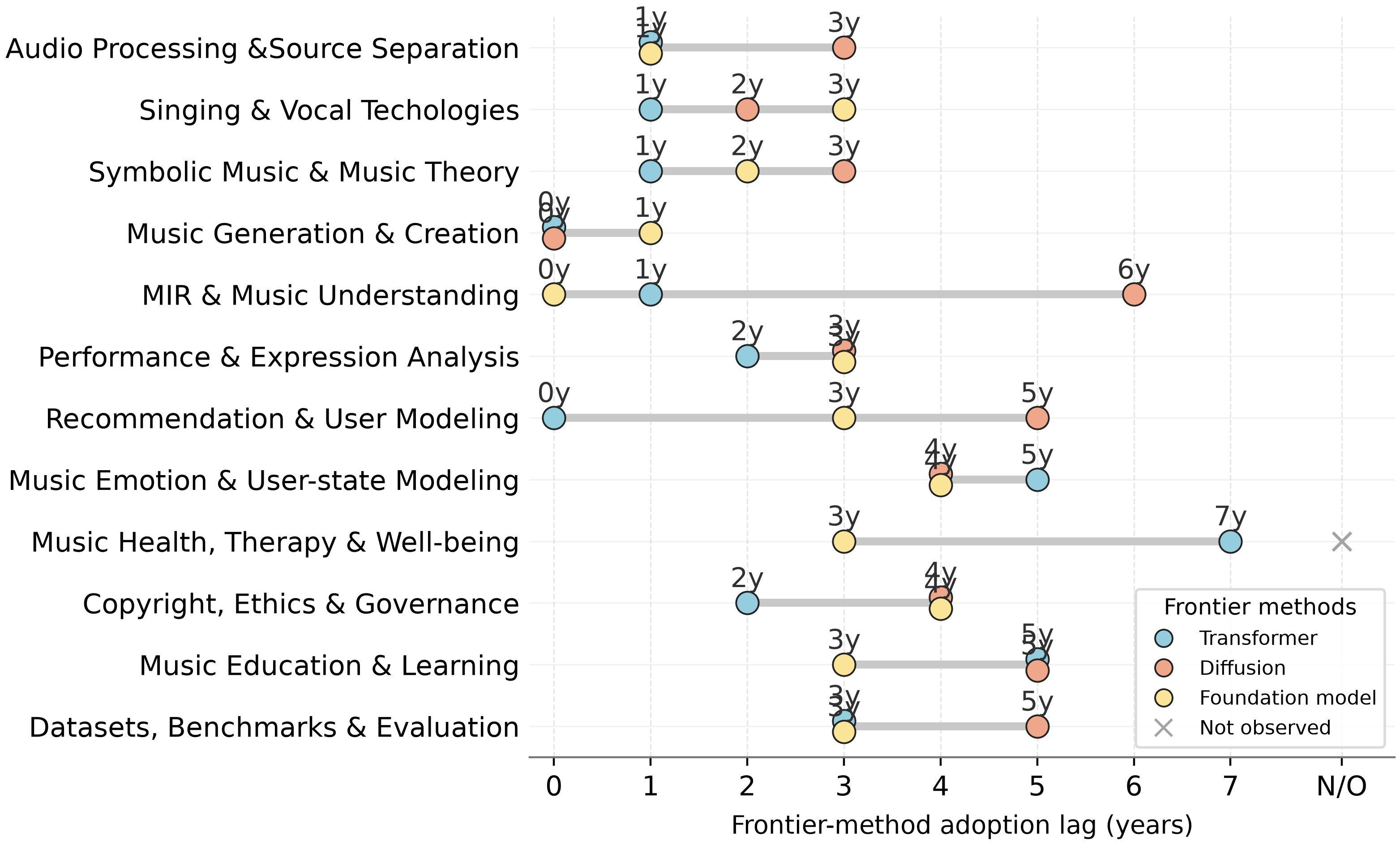}
    \caption{Average Frontier Method Adoption Lag (FMAL) across AI music application categories. Lower values indicate earlier adoption of frontier methods, whereas higher values indicate delayed adoption.}
    \label{fig:FMAL}
\end{figure}

Fig.~\ref{fig:FMAL} further reveals substantial differences in the timing of frontier-method adoption. Generation adopted the three frontier-method families with an average lag of only $0.33$ years, whereas MIR showed a mean lag of $2.33$ years. Education and governance lagged by $4.33$ and $3.33$ years, respectively. Health showed the largest mean lag among adopted methods ($5.00$ years), while no diffusion-based study was observed during the study period.
\section{Discussion}

\paragraph{Frontier methods are selectively concentrated.}
The growth of AI music research has not led to a proportional distribution of methodological investment across application areas. Audio processing, singing, and generation received more technical investment than expected from their publication volumes
($\mathrm{TIR}=0.331$, $0.299$, and $0.134$), whereas education, governance, and health received less
($-0.466$, $-0.404$, and $-0.282$). This imbalance becomes more pronounced when frontier methods are considered. Generation attracted substantially more Transformer- and diffusion-based investment than expected
($\mathrm{TMIR}=0.902$ and $1.317$), while foundation-model investment was particularly concentrated in MIR
($\mathrm{TMIR}=0.973$). In contrast, education, health, and governance showed negative residuals across Transformer, diffusion, and foundation-model families.

These patterns suggest that frontier methods do not diffuse uniformly across the field, but are absorbed first by tasks with scalable datasets, standardized objectives, established benchmarks, and a close fit between model capabilities and task formulation. Generation and audio-centered tasks satisfy many of these conditions, whereas education, health, and governance often require contextual evaluation, interdisciplinary expertise, longitudinal evidence, or stronger ethical and regulatory safeguards. The observed concentration therefore appears to reflect not only differences in research popularity, but also unequal task-level capacity to absorb emerging methods. 

\paragraph{Methodological progress favors tasks supported by scalable data infrastructures.}
The concentration of frontier methods may partly reflect differences in the computational scalability of application tasks. Generation, MIR, and audio processing are supported by relatively reusable datasets, standardized metrics, and repeatable benchmark pipelines, making it easier to train, compare, and extend new model architectures. This infrastructure has also expanded rapidly: dataset- and benchmark-related publications increased from 8 in 2015 to 56 in 2022 and 144 in 2025, with 322 papers published during 2023--2025, compared with 169 during 2018--2022 and 34 during 2015--2017. The accumulation of datasets, annotations, and evaluation resources may therefore have created favorable conditions. By contrast, education, health, and governance often require context-specific data, institutional participation, longitudinal outcomes, and domain-sensitive evaluation, making them less readily standardized and scaled. Although these temporal patterns do not establish causality, they suggest that methodological progress is shaped not only by model innovation, but also by whether a task has the data and evaluation infrastructure needed to absorb it.

\paragraph{Socially embedded applications receive frontier methods later.}
The temporal results further indicate that frontier methods diffuse unevenly across application domains. Generation adopted the three frontier-method families with an average lag of only $0.33$ years, compared with $4.33$ years for education and $5.00$ years for health among adopted methods; no diffusion-based health study was observed during the study period. This contrast  may reflect a model-driven situation: Transformer, diffusion, and foundation models can be more readily translated into generation-oriented tasks, where their capabilities are directly expressed through content production and evaluated with established datasets and benchmarks. Education, health, and governance, by contrast, often require domain-specific validation, contextual interpretation, interdisciplinary expertise, and evidence of real-world effectiveness before new methods can be meaningfully adopted.

Methodological diversity alone did not compensate for limited investment. Governance had a relatively high NMD ($0.864$), but this estimate was based on only 39 method-mapped papers, indicating diversity within a small technical literature rather than broad methodological support. Taken together, these findings suggest that the diffusion of frontier methods may be shaped more strongly by compatibility with existing model capabilities and research infrastructures than by the relative societal importance of different application domains.

\paragraph{Toward a more socially responsive research agenda.}
The observed imbalance raises an alignment question: whether the rapid expansion of generative capability is being matched by comparable methodological support for socially embedded applications. While generation received above-expected technical investment and adopted frontier methods rapidly, education, health, and governance remained below expectation and showed substantially longer adoption lags. This pattern does not imply lower societal relevance, but suggests that current research infrastructures are better suited to tasks with scalable datasets, standardized objectives, and readily measurable outputs. A more socially responsive agenda therefore requires not only additional model development, but also task-appropriate datasets, context-sensitive evaluation, longitudinal validation, and stronger collaboration across AI, music, education, health, and governance communities.

\section{Limitations}
Our corpus does not cover all relevant publications, and the 2026 data are incomplete because collection ended in April. Assigning one primary task and method also simplifies cross-task and multi-method studies, while first-adoption years may be sensitive to a few early papers. More importantly, our indicators measure research attention rather than societal demand or realized impact. Future work could incorporate citations, funding, collaboration, dataset dependencies, and real-world deployment.
\section{Conclusion}

This study analyzed 6,839 AI music publications and introduced the Research Attention Profile framework to measure how technical attention is distributed across tasks. The results reveal persistent imbalances: generation, MIR, and audio processing attract greater technical investment and adopt frontier methods earlier, while education, health, and governance remain comparatively limited technical investment. Future work will examine how these patterns relate to relevant Sustainable Development Goals and cultural-sustainability priorities, supporting a more balanced development of AI music research.

\section*{Data Availability}
The dataset constructed in this study is available from the corresponding author upon reasonable request.


\bibliographystyle{unsrtnat}
\bibliography{references}

\clearpage
\includepdf[pages=-]{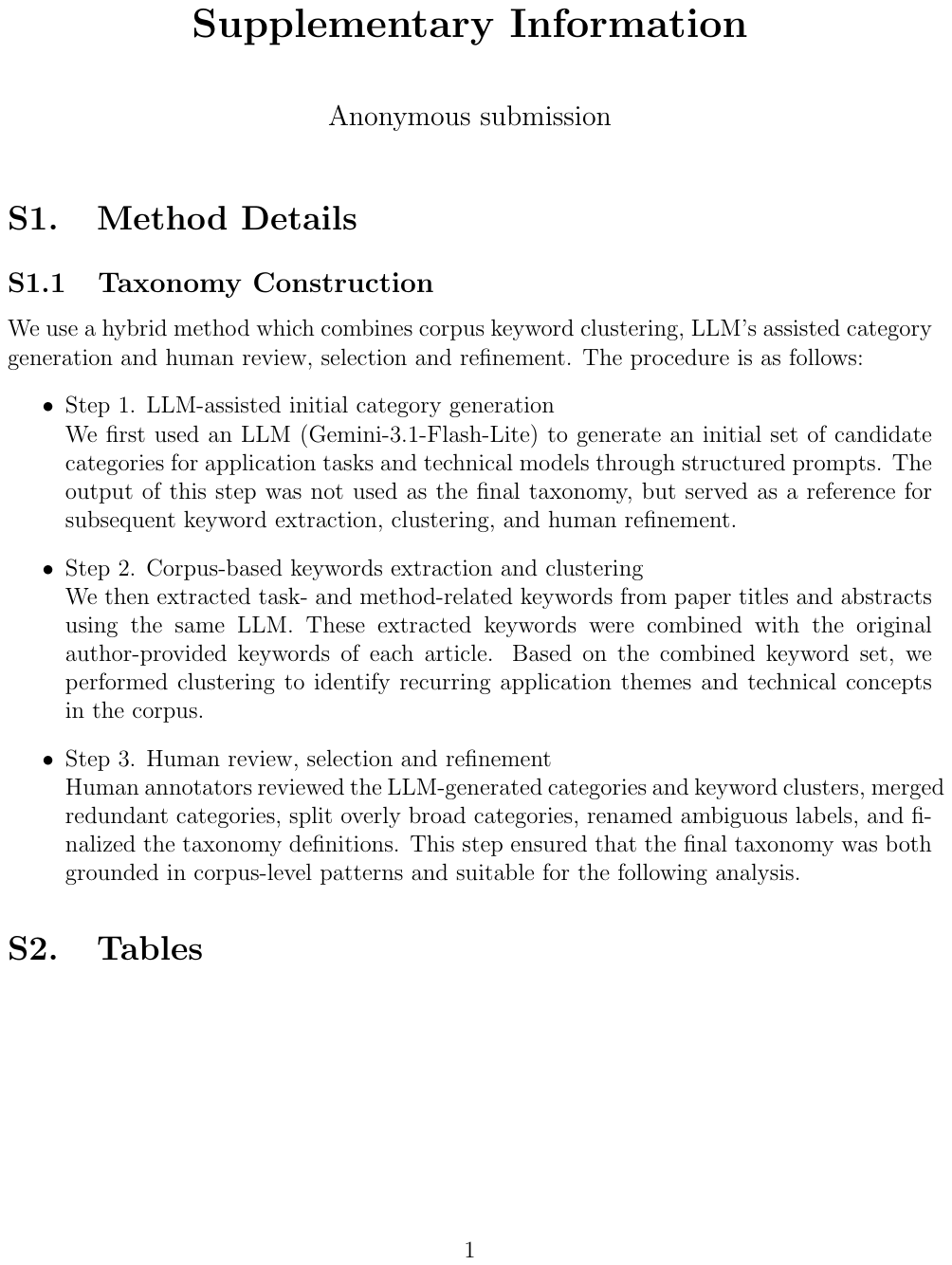}

\end{document}